\documentclass[12pt]{article}

\usepackage{amssymb}

\begin{document}

\title{Lifshitz field theories, Snyder noncomutative  space-time and  momentum dependent metric  }

\author{ Juan M. Romero\thanks{jromero@correo.cua.uam.mx}\\[0.5cm]
\it Departamento de Matem\'aticas Aplicadas y Sistemas,\\
\it Universidad Aut\'onoma Metropolitana-Cuajimalpa,\\
\it M\'exico, D.F  05300, M\'exico\\[0.5cm]\\
 J. David Vergara\thanks{vergara@nucleares.unam.mx}
\\[0.5cm]
\it Instituto de Ciencias Nucleares,\\
\it Universidad Nacional Aut«onoma de M\'exico,\\
\it A. Postal 70-543 , M\'exico D.F., M\'exico\\
} 

\date{}

\maketitle

\pagestyle{plain}

\begin{abstract}

In this work, we propose three different modified relativistic particles. 
In the first case, we propose a particle with metrics depending  on the momenta and 
we show that the quantum version of these systems includes different field theories, as Lifshitz  field theories. 
As a second case we propose a particle that implies a modified symplectic structure and we show that the quantum version of this system gives different noncommutative space-times, for example the Snyder space-time. In the third  case, we combine both structures  before mentioned, namely noncommutative space-times and momentum dependent metrics. In this last case, we show that anisotropic field theories can be seen as a limit of noncommutative field theory. 
\end{abstract}

\section{Introduction}

Recently, different approaches have been developed to obtain a quantum version of gravity.  Some of these approaches  are  string theory \cite{polchinski:gnus},  loop quantum gravity \cite{rovelli:gnus}, noncommutative geometry \cite{connes:gnus}, etc.  
In $(2+1)$ dimensions there are important progress \cite{carlip:gnus,w1:gnus}, but in $(3+1)$ dimensions we do not know how this theory  is, we only have   some signs about it.  For instance, using the Ehrenfest principle, Bekenstein proposed that in a quantum gravity  the area of the event horizon has discrete spectrum \cite{beke1:gnus,beke2:gnus}
 \begin{eqnarray}
 A_{n}=4\pi r^{2}=\gamma l_{p}^{2}n, \qquad n=1,2,\cdots. 
 \end{eqnarray}
In addition, G.  `t Hooft showed that  in $(2+1)$ dimensions the quantum gravity implies a discrete space-time in an effective approximation  \cite{hooft:gnus}. 
For those reasons, we can conjecture that in the quantum gravity there are geometric quantities with discrete spectrum. Remarkably, the discrete space-time  obtained by  G. `t Hooft  is the so-called Snyder space-time, which is discrete, noncommutative and compatible with the Lorentz symmetry \cite{s:gnus}.  In fact, in this noncommutative space-time the surface  area of a sphere is quantized \cite{za:gnus}.   It is worth mentioning that it is possible construct field theories in some noncommutative space-times \cite{w2:gnus,douglas:gnus,szabo:gnus}, but to build a gravity  or field theory in a noncommutative space-time, as Snyder space-time, is a very difficult task. Some work about Snyder space-time can be seen in \cite{snyder1:gnus,snyder2:gnus,snyder3:gnus,snyder4:gnus,snyder5:gnus,snyder6:gnus,snyder7:gnus,snyder8:gnus,snyder9:gnus,snyder10:gnus,snyder11:gnus,snyder12:gnus,snyder13:gnus,snyder14:gnus} and  works about  noncommutative space-times which imply discrete geometric quantities can be seen in \cite{dis1:gnus,dis2:gnus,dis3:gnus,dis4:gnus,dis5:gnus}. \\

A major problem to obtain a quantum gravity theory is that the usual gravity is unrenormalizable. Nevertheless,  recently  Ho\v{r}ava formulated a modified gravity which seems to be free ghosts and power counting renormalizable  \cite{Horava:gnus}. This gravity is invariant under anisotropic scaling
\begin{eqnarray}
\vec x \to b \vec x, \qquad t \to b^{z}t, \quad z, b={\rm constants} ,   \label{eq:scaling}
\end{eqnarray}
with  $z=3.$ The original  Ho\v{r}ava gravity has dynamical inconsistencies \cite{Henneaux:gnus}, but   were found healthy extensions of it \cite{Pujolas:gnus,pujolas2:gnus}. 
Ho\v{r}ava gravity has different interesting properties, some of them   are  
\cite{horava1:gnus,horava2:gnus,horava3:gnus,horava4:gnus,horava5:gnus,horava6:gnus,horava7:gnus,horava8:gnus,horava9:gnus}. 
Field theories invariant  under the anisotropic scaling transformations 
 (\ref{eq:scaling}) can be seen in 
 \cite{anselmi:gnus,anselmi1:gnus,anselmi2:gnus,anselmi3:gnus,anselmi4:gnus,anselmi5:gnus,anselmi6:gnus,anselmi7:gnus,anselmi8:gnus,anselmi9:gnus,anselmi10:gnus}, notably  these field theories
 improve their  high energy behavior. Furthermore, in the usual  general relativity it has been found  space-times invariant under the anisotropic scaling (\ref{eq:scaling}), see \cite{marika:gnus}. Significantly, using the gravity/gauge correspondence,  these space-times  can be related with some condensed matter systems \cite{marika5:gnus,marika6:gnus,marika7:gnus}.\\

Now, as a  road to obtain new physics,  different authors have been proposed that the Minkowski geometry should be changed by the Finsler geometry \cite{finsler1:gnus,finsler2:gnus,finsler3:gnus,finsler4:gnus,finsler5:gnus,finsler6:gnus}. In this new geometry  the metric depends on velocities, notice that in this case the metric can depends  on the momenta.   Remarkably,  since 1938 Max Born proposed a theory with a metric that depends on the momenta as a suggestion for unifying quantum theory and relativity \cite{born:gnus}.\\
  
In this work, we propose three different  modified relativistic particles.  In the first case, we  propose a particle with a metric depending  on the momenta and 
we show that the quantum version of this system includes  different field theories,  as anisotropic field theories. As a second case we propose a particle  that implies a modified symplectic structure and we show  that the quantum version of this system gives  different noncommutative space-times, for example the Snyder space-time. In the third case, we combine both structures  before mentioned, namely  noncommutative space-times and momentum-dependent metric. In this last case, we show that anisotropic field theories can be seen as a limit of noncommutative field theory. \\

 This  paper is organized as follow: in  Sec. 2,  we study the first modified relativistic particle and show that in this framework different Lifshitz field theories can be obtained;
 in Sec. 3, we  propose a modified particle that its quantum version implies noncommutative space-times; in Sec.  4, we combine the results obtained in Sec. 2 and 3. Finally, in Sec.  5, we provide a summary.

\section{Modified actions}

The Hamiltonian action for the massive relativistic particle with the momenta fixed in the end points is given by
\begin{eqnarray}
S=\int d\tau\left(  - \dot p_{\mu} x^{\mu} - \frac{\lambda}{2} \left(p^{2}-m^{2} \right)\right)= 
\int d\tau \left(  -\eta ^{\mu\nu}\dot p_{\mu} x_{\nu} - \frac{\lambda}{2} \left(\eta^{\mu\nu} p_{\mu}p_{\nu} -m^{2} \right) \right),\quad 
\end{eqnarray}
where $\lambda$ is a Lagrange multiplier.  Now, if we consider a momentum-dependent metric
\begin{eqnarray}
\Omega^{\mu \nu}(p),
\end{eqnarray}
we can  propose the generalized relativistic particle action
\begin{eqnarray}
S= \int d\tau \left(  -\eta ^{\mu\nu}\dot p_{\mu} x_{\nu} - \frac{\lambda}{2} \left(\Omega^{\mu\nu} (p)p_{\mu}p_{\nu} -m^{2} \right) \right). \label{eq:a}
\end{eqnarray}
In the section 3, we will show that the quantum version of this system includes different field theories,  as Lifshitz  field theories. \\

In addition,  if 
\begin{eqnarray}
\Lambda ^{\mu \nu}(p)
\end{eqnarray}
is a symmetric matrix we can also introduce the  alternative  action
\begin{eqnarray}
S=\int d\tau \left( - \Lambda^{\mu \nu}(p)  \dot p_{\mu} x_{\nu} - \frac{\lambda}{2} \left(\eta^{\mu\nu} p_{\mu}p_{\nu} -m^{2} \right)   \right). \label{eq:b}
\end{eqnarray}
Notice that in this case the space-time metric is not modified. In the section 4, we will show that this system has a modified symplectic structure and 
also  we will show  that the quantum version of this system gives  different noncommutative space-times, for example the Snyder space-time. \\

Furthermore, we can take the action
\begin{eqnarray}
S=\int d\tau \left( - \Lambda^{\mu \nu}(p)  \dot p_{\mu} x_{\nu} - \frac{\lambda}{2} \left(\Omega^{\mu\nu}(p) p_{\mu}p_{\nu} -m^{2} \right)   \right), \label{eq:c}
\end{eqnarray}
which contains  the actions (\ref{eq:a}) and (\ref{eq:b}). In all these cases, we assume that we use the Minkowski metric for  raising and lowering indices. \\
Notice that these three  actions are  invariant under  reparametrization  transformations
\begin{eqnarray}
\tau \to \tau(\tilde \tau) ,\qquad \lambda \to \lambda\frac{d\tilde \tau}{d\tau}, \nonumber 
\end{eqnarray}
in fact this symmetry appears in the usual relativistic particle \cite{ct:gnus}.  Due that the reparametrization  symmetry is a local symmetry, 
according to Dirac's Method \cite{dirac1:gnus,ct:gnus} these three action have a first  class constraint
\begin{eqnarray}
C(x,p)\approx 0, \nonumber
\end{eqnarray}
which generates the  reparametrization symmetry (the  "gauge  symmetry"  for these systems). Now, if $A(x,p)$ is a function in the phase space,  according to the Dirac's   Method, an infinitesimal   gauge transformation are given by 
\begin{eqnarray}
\delta A= \epsilon (x,p)\{A(x,p),C(x,p)\}\nonumber .
\end{eqnarray}
Notice that only if $A(x,p)$ is a gauge invariant quantity we have 
\begin{eqnarray}
\{A(x,p),C(x,p)\}=0. \nonumber
\end{eqnarray}
Furthermore, at quantum level Dirac's method   sets  that  the physical states  are invariant under the action of the first class constraints, i.e.,
\begin{equation}
\exp(\zeta \hat C)|\psi\rangle=|\psi\rangle ,
\end{equation} 
which implies
\begin{equation}
\hat C|\psi\rangle=0.
\end{equation} 
Here $\hat C$ is the quantum version of the constraint $C(x,p).$\\

In the next sections we will show that these three  actions are different and each  one of them gives an alternative  quantum physics.

\section{Lifshitz case }

First we consider the following  action
\begin{eqnarray}
S= \int d\tau \left(  -\eta ^{\mu\nu}\dot p_{\mu} x_{\nu} - \frac{\lambda}{2} \left(\Omega^{\mu\nu} (p)p_{\mu}p_{\nu} -m^{2} \right) \right),
\end{eqnarray}
from this  action we obtain the classical constraint
\begin{eqnarray}
C= \Omega^{\mu\nu}(p)  p_{\mu} p_{\nu} -m^{2}\approx 0.
\end{eqnarray}
Then, using the Dirac's Method,  at quantum level and in the coordinate representation we get the wave equation 
\begin{eqnarray}
\left ( -\Omega^{\mu\nu}(-i \partial )  \partial _{\mu} \partial _{\nu} -m^{2}\right) \phi =0, \label{eq:mkg}
\end{eqnarray}
which  is a modified Klein-Gordon equation. \\

\subsection{Scalar Field }

The equation (\ref{eq:mkg}) can be obtained from the action
\begin{eqnarray}
S=\int dx^{d+1} \frac{1}{2}\left(\partial_{\mu}\phi\Omega^{\mu\nu}\left(-i \partial  \right) \partial_{\nu}\phi-m^{2}  \phi^{2} \right).
\end{eqnarray}
In fact, if the matrix $\Omega^{\mu\nu}\left(-i\partial\right)$ has only an even number of derivatives, we  arrive to 

\begin{eqnarray}
\delta S=-\int dx^{d+1} \delta \phi \left( \Omega^{\mu\nu}\left(-i\partial\right) \partial_{\mu}\partial_{\nu} \phi+m^{2} \phi  \right)=0,
\end{eqnarray}
which implies the equation of motion 
\begin{eqnarray}
\Omega^{\mu\nu}\left(-i\partial \right) \partial_{\mu}\partial_{\nu} \phi+m^{2} \phi=0,
\end{eqnarray}
that is equivalent to the equation (\ref{eq:mkg}).\\

Notice, that if
\begin{eqnarray}
\Omega^{\mu\nu}\left(-i\partial \right)= \eta^{\mu\nu}+h^{\mu\nu}\left(-i\partial \right), 
\end{eqnarray}
we have 
\begin{eqnarray}
S&=&\int dx^{d+1} \frac{1}{2}\left(\partial_{\mu} \phi \partial^{\mu}\phi+\left( \partial_{\mu} \phi \right)h^{\mu\nu}\left(-i\partial \right)\left(\partial_{\nu} \phi \right) -m^{2} \phi^{2} \right)\nonumber\\
&=& \int dx^{d+1} \frac{1}{2}\left(\partial_{\mu} \phi \partial^{\mu}\phi-  \phi h^{\mu\nu}\left(-i\partial \right)\left( \partial_{\mu}\partial_{\nu} \phi \right) -m^{2} \phi^{2}  \right).\label{eq:scalar}
\end{eqnarray}
In particular, if we take
\begin{eqnarray}
\Omega^{\mu\nu}(p)= \eta^{\mu\nu}+l^{2}p^{\mu} p^{\nu}, \qquad l={\rm constant},
\end{eqnarray}
namely
\begin{eqnarray}
\Omega^{\mu\nu}\left( -i \partial \right)= \eta^{\mu\nu}+l^{2}\hat p^{\mu} \hat p^{\nu}= \eta^{\mu\nu}- l^{2} \partial^{\mu}\partial^{\nu},
\end{eqnarray}
 we arrive to 
\begin{eqnarray}
S&=& \int dx^{d+1} \frac{1}{2}\left(\partial_{\mu} \phi \partial^{\mu}\phi +l^{2} \phi \Box ^{2}
\phi -m^{2}  \phi^{2} \right).
\end{eqnarray}
In general, if $G(p^{2})$ is a smooth function, when 
\begin{eqnarray}
\Omega^{\mu\nu}(p)= \eta^{\mu\nu}+ G\left(p^{2}\right)p^{\mu} p^{\nu},
\end{eqnarray}
we obtain 
\begin{eqnarray}
\Omega^{\mu\nu}\left( -i\partial\right)= \eta^{\mu\nu} - G\left( -\Box  \right)  \partial^{\mu}\partial^{\nu},
\end{eqnarray}
for this case, the action (\ref{eq:scalar}) becomes 
\begin{eqnarray}
S&=& \int dx^{d+1} \frac{1}{2}\left(\partial_{\mu} \phi \partial^{\mu}\phi +\phi G\left( -\Box \right) \Box ^{2}
\phi -m^{2} \phi ^{2} \right).
\end{eqnarray}
This  is a quantum field theory with high order time derivatives, which implies the existence of ghost field solutions 
\cite{Bernard}. To avoid these kind of solutions  
we introduce the matrix
\begin{eqnarray}
\Omega^{0\mu}(p)=\eta^{0\mu}, \qquad  \Omega^{ij}(p)= \delta^{ij} + l^{2}p^{i}p^{j},
\end{eqnarray}
the quantum version of this last equation is given by
\begin{eqnarray}
\Omega^{0\nu}\left(-i\partial \right)=\eta^{0\mu}, \qquad  \Omega^{ij}(p)= \delta^{ij}-l^{2}\partial^{i}\partial^{j}, 
\end{eqnarray}
and for the action (\ref{eq:scalar}) we get
\begin{eqnarray}
S&=&\int dx^{d+1} \frac{1}{2}\left(\partial_{\mu} \phi \partial^{\mu}\phi- l^{2}\partial_{i} \phi \partial^{i}\partial^{j} \partial_{j}\phi -m^{2}  \phi^{2} \right)\nonumber\\
 &=&\int dx^{d+1} \frac{1}{2}\left(\partial_{\mu} \phi \partial^{\mu}\phi+l^{2} \phi \left(\nabla^{2}\right)^{2}\phi -m^{2} \phi^{2} \right).
\end{eqnarray}
This is  the action for an  anisotropic scalar field with dynamic exponent $z=2$, also this field is named Lifshitz  scalar field \cite{anselmi6:gnus,anselmi10:gnus}. \\

Moreover,   for the case 
\begin{eqnarray}
\Omega^{0\nu}(p)=\eta^{0\nu}, \qquad \Omega^{ij}(p)= \delta^{ij} +\alpha \left(- \vec p^{\ 2} \right)^{z-2}p^{i}p^{j},
\end{eqnarray}
namely
\begin{eqnarray}
\Omega^{0\nu}(-i\partial )=\eta^{0\nu}, \qquad  \Omega^{ij}(-i\partial )= \delta^{ij} -  \alpha \left(  \nabla^{2}\right)^{z-2}\partial ^{i}\partial ^{j},
\end{eqnarray}
the  action (\ref{eq:scalar}) becomes
\begin{eqnarray}
S&=&\int dx^{d+1} \frac{1}{2}\left(\partial_{\mu} \phi \partial^{\mu}\phi+ \alpha \phi \left( \nabla^{2} \right)^{z} \phi -m^{2}  \phi^{2} \right).
\label{eq:lif}
\end{eqnarray}
Which is the action for an anisotropic scalar field with dynamic exponent $z,$ \cite{anselmi6:gnus,anselmi10:gnus}.
Then, a Lifshitz scalar field can be seen as a scalar field in a generalized  metric depending  on the momenta. \\

In general, if $G(\vec p^{\ 2})$ is a smooth function, we can propose the matrix  
\begin{eqnarray}
\Omega^{0\nu}(p)=\eta^{0\nu}, \qquad \Omega^{ij}(p)= \delta^{ij} +G \left( \vec p^{\ 2} \right) p^{i}p^{j},
\end{eqnarray}
which at quantum level is 
\begin{eqnarray}
\Omega^{0\nu}\left(-i\partial \right)=\eta^{0\nu}, \qquad \Omega^{ij}\left(-i\partial \right)= \eta^{ij}-G\left( -\nabla^{2} \right)\partial^{i}\partial^{j},
\end{eqnarray}
gives the action 
\begin{eqnarray}
S&=&\int dx^{d+1} \frac{1}{2}\left(\partial_{\mu} \phi \partial^{\mu}\phi+  \phi G\left( -\nabla^{2} \right)\left(\nabla^{2}\right)^{2}\phi -m^{2} \phi^{2}  \right).
\end{eqnarray}
In the next subsections we will study other fields in a momentum dependent metric.

\subsection{Dirac Field }

Now we construct a Dirac equation in a metric that depends on the momenta. 
In this case, we require  a tetrad formalism associate to the momenta dependent metric. 
For this reason, let us introduce the tetrad 
\begin{eqnarray}
e_{a}^{\mu}=  e_{a}^{\mu}\left(-i \partial \right),  \label{eq:tetrad}
\end{eqnarray}
which satisfies 
\begin{eqnarray}
e_{a}^{\mu}\left(-i \partial \right)e_{b}^{\nu}\left(-i \partial \right)\eta^{ab}= \Omega^{\mu\nu}\left(-i \partial \right).
\end{eqnarray}
Then, using the usual Dirac's matrices,  that satisfy the ordinary Clifford algebra 
\begin{eqnarray}
\left \{\gamma^{a}, \gamma^{b}\right\}_{+} =2\eta^{ab}, 
\end{eqnarray}
and the tetrad basis introduced in (\ref{eq:tetrad}) we obtain the following  matrices 
\begin{eqnarray}
\Gamma^{\mu}\left(-i \partial \right)= e_{a}^{\mu} \left(-i \partial \right)\gamma^{a},
\end{eqnarray}
which satisfy
\begin{eqnarray}
\left \{\Gamma^{\mu}\left(-i \partial \right), \Gamma^{\nu}\left(-i \partial \right)\right\}_{+} =2\Omega^{\mu\nu}\left(-i \partial \right). \label{eq:dm}
\end{eqnarray}
With the matrices $\Gamma^{\mu}$ we propose the modified Dirac equation 
\begin{eqnarray}
-i\Gamma^{\mu}\partial_{\mu} \psi +m\psi=0.
\end{eqnarray}
Notice that  using this last equation and (\ref{eq:dm}), we arrive to
\begin{eqnarray}
\left(- i\Gamma^{\nu}\partial_{\nu}-m\right)  \left(- i\Gamma^{\mu}\partial_{\mu} +m\right)\psi &=&\nonumber\\
&= & -\Gamma^{\mu}\Gamma^{\nu} \partial_{\mu}\partial_{\nu}\psi -m^2\psi\nonumber\\
& =&\left( -\frac{\Gamma^{\mu}\Gamma^{\nu} + \Gamma^{\nu}\Gamma^{\mu} }{2}   \partial_{\mu}\partial_{\nu}-m^{2}\right)\psi=0,\nonumber
\end{eqnarray}
namely 
\begin{eqnarray}
-\left( \Omega^{\mu\nu} \left(-i \partial \right) \partial_{\mu}\partial_{\nu}+m^{2}\right)\psi=0,
\end{eqnarray}
which is the modified Klein-Gordon  equation (\ref{eq:mkg}).
Then, the generalized  Dirac's equation can be seen as a  Dirac's equation in a metric depending  on the momenta. \\

In particular,  if we take 
\begin{eqnarray}
\Omega^{\mu\nu}\left(-i \partial \right)=\eta^{\mu\nu}+ h^{\mu\nu}\left(-i \partial \right),
\end{eqnarray}
 at first order, the tetrad results 
\begin{eqnarray}
e_{a}^{\mu}\left(-i \partial \right)= \eta_{a}^{\mu}+\frac{1}{2} h^{\mu}_{a} \left(-i \partial \right),
\end{eqnarray}
which satisfies 
\begin{eqnarray}
e_{a}^{\mu}e_{b}^{\nu}\eta^{ab}\approx  \eta^{\mu\nu}+ h^{\mu\nu}\left(-i \partial \right).
\end{eqnarray}
In this approximation  we obtain 
\begin{eqnarray}
\Gamma^{\mu}=  \gamma^{\mu}+\frac{1}{2} h^{\mu}_{a}\gamma^{a}
\end{eqnarray}
and the modified Dirac's equation is given by 
\begin{eqnarray}
\left(-i \gamma^{\mu}\partial_{\mu} +m\right)\psi -i \frac{1}{2} h^{\mu}_{a}\gamma^{a} \partial_{\mu} \psi =0.
\end{eqnarray}
For the case 
\begin{eqnarray}
h^{\mu\nu}\left(-i \partial \right)  = - G\left(-\Box\right) \partial^{\mu}\partial^{\nu}, 
\end{eqnarray}
we arrive to 
\begin{eqnarray}
\left( -i\gamma^{\mu}\partial_{\mu} +m \right)\psi+\frac{i}{2} G\left(-\Box\right)  \partial^{\mu} \partial_{a} \gamma^{a} \partial_{\mu} \psi=0,
\end{eqnarray}
namely
\begin{eqnarray}
\left(-i \gamma^{\mu} \partial_{\mu}+m\right) \psi+ \frac{i}{2} G\left(-\Box\right)  \Box  \gamma^{\mu} \partial_{\mu}\psi=0. 
\end{eqnarray}
While, if we take 
\begin{eqnarray}
h^{\mu 0 }\left(-i \partial \right) =0,\qquad h^{ij}\left(-i \partial \right)  = - G\left(-\nabla^{2} \right) \partial^{i}\partial^{j}, 
\end{eqnarray}
we get 
\begin{eqnarray}
\left(-i \gamma^{\mu}  \partial_{\mu}+m \right)\psi+  \frac{i}{2} G\left(-\nabla^{2} \right)  \nabla^{2}  \gamma^{\mu}  \partial_{\mu}\psi=0. 
\end{eqnarray}
In particular when 
\begin{eqnarray}
G\left( -\nabla^{2} \right)= \alpha \left(\nabla^{2}\right)^{z-2},\qquad \alpha={\rm constant},
\end{eqnarray}
we obtain
\begin{eqnarray}
\left(-i \gamma^{\mu}  \partial_{\mu}+m \right)\psi+  \frac{i}{2} \alpha \left(-\nabla^{2} \right)^{z-1}   \gamma^{\mu}  \partial_{\mu}\psi=0,
\end{eqnarray}
which is the anisotropic Dirac's equation with dynamic exponent $z,$ \cite{anselmi:gnus,anselmi10:gnus}.
Therefore,  the anisotropic Dirac's equation  can be seen as a Dirac's equation in a metric depending on the momenta. \\

\subsection{Electromagnetic Field }

For the electromagnetic field we propose the action
\begin{eqnarray}
S=-\frac{1}{4} \int dx^{d+1}F_{\mu\nu} \Omega^{\mu\alpha}\left(-i \partial  \right) \Omega^{\nu\beta}\left(-i \partial  \right) F_{\alpha \beta}.
\end{eqnarray}
Notice that if 
\begin{eqnarray}
\Omega^{\mu\nu}\left(-i\partial \right)= \eta^{\mu\nu}+h^{\mu\nu} \left(-i\partial \right),
\end{eqnarray}
at first order,  we obtain  
\begin{eqnarray}
\Omega^{\mu\alpha}\left(-i\partial \right)\Omega^{\nu\beta}\left(-i\partial \right)&=& \left(\eta^{\mu\alpha}+h^{\mu\alpha}\right) \left(\eta^{\nu\beta}+h^{\nu\beta}\right)\nonumber\\ 
&\approx & \eta^{\mu\alpha}\eta^{\nu\beta}+h^{\mu\alpha}\left(-i\partial \right) \eta^{\mu\beta}+\eta^{\mu\alpha}h^{\nu\beta}\left(-i\partial \right),
\end{eqnarray}
which  implies 
\begin{eqnarray}
F_{\mu\nu} \Omega^{\mu\alpha}\left(-i\partial \right) \Omega^{\nu\beta} \left(-i\partial \right)F_{\alpha \beta}\approx F_{\mu\nu}F^{\mu\nu} +
2F_{\mu}\-^{\beta}h^{\mu\alpha}\left(-i\partial \right) F_{\alpha\beta}.
\end{eqnarray}
Moreover, using this last equation we arrive to 
\begin{eqnarray}
S=-\frac{1}{4}\int dx^{d+1}\left(F_{\mu\nu}F^{\mu\nu} +
2F_{\mu}\-^{\beta}h^{\mu\alpha}\left(-i\partial \right)F_{\alpha\beta}\right). 
\end{eqnarray}
In particular, when  
\begin{eqnarray}
h^{\mu\nu}\left(-i \partial \right)  = - G\left(-\Box\right) \partial^{\mu}\partial^{\nu}, 
\end{eqnarray}
 we get the action 
\begin{eqnarray}
S&=&-\frac{1}{4}\int dx^{d+1}\left(F_{\mu\nu}F^{\mu\nu} 
-2F_{\mu}\-^{\beta}G\left(-\Box\right)\partial ^{\mu}\partial^{\alpha}F_{\alpha\beta}\right)\nonumber\\
&=&\int dx^{d+1}\left(F_{\mu\nu}F^{\mu\nu} 
+2\partial^{\mu} F_{\mu}\-^{\beta}G\left(-\Box\right)\partial^{\alpha}F_{\alpha\beta}\right).
\end{eqnarray}
Furthermore, if $h^{\mu\nu}$ is given by
\begin{eqnarray}
h^{\mu 0 }\left(-i \partial \right) =0,\qquad h^{ij}\left(-i \partial \right)  = - G\left(-\nabla^{2} \right) \partial^{i}\partial^{j}, 
\end{eqnarray}
the following action 
\begin{eqnarray}
S&=&-\frac{1}{4}\int dx^{d+1}\left(F_{\mu\nu}F^{\mu\nu} 
+2\partial_{i} F_{ik}G\left(-\nabla^{2} \right)\partial_{j}F_{jk}\right)
\end{eqnarray}
is obtained.\\

Notice that,  when $G\left(-\nabla^{2}\right)=l^{2},$   we arrive to the action
\begin{eqnarray}
S= -\frac{1}{4}\int dx^{d+1}\left(F_{\mu\nu}F^{\mu\nu} +
2l^{2} \partial_{i}F_{ik} \partial_{j}F_{jk} \right),
\end{eqnarray}
which is  the action for the anisotropic electrodynamics field  with dynamic exponent  $z=2,$ \cite{anselmi2:gnus}.
Then,  this last system can be seen as an  electrodynamics in a metric depending on the momenta. 

\subsection{Yang-Mills Field }

It is well known that for  non abelian fields the usual derivative is changed by the covariant derivative, in the sense that the partial derivative is not gauge covariant. For this reason, for non abelian gauge fields  the matrix $\Omega^{\mu\nu}$ is changed by 
\begin{eqnarray}
 \Omega^{\mu\nu}(-i\partial)  \to \Omega^{\mu\alpha}\left(-i D \right),   
\end{eqnarray}
where 
\begin{eqnarray}
D_{\mu}{\cal F}=\partial_\mu {\cal F} +[{\cal A}_\mu,{\cal F}],
\end{eqnarray}
for ${\cal F}$ in a matrix representation of the Lie group.
Hence, in this framework the usual Yang-Mills action is changed  by
\begin{eqnarray}
S=\frac{1}{2g^2}\int dx^{d+1} Tr \left( 
 {\cal  F}_{\mu\nu} \Omega^{\mu\alpha}\left(-i D \right) \Omega^{\nu\beta}\left(-i D  \right) {\cal F}_{\alpha \beta} \right).
\end{eqnarray}
which is  gauge invariant. In particular  when
\begin{eqnarray}
\Omega^{\mu\nu}= \eta^{\mu\nu}+h^{\mu\nu}\left(-i D \right),
\end{eqnarray}
at  first order, we get 
\begin{eqnarray}
S=\frac{1}{2g^{2}}\int dx^{d+1} Tr \left( 
 {\cal  F}_{\mu\nu}  {\cal F}^{\mu \nu}+ 2 {\cal F}_{\mu}\-^{\beta} h^{\mu\alpha}\left(-i D \right) {\cal F}_{\alpha \beta}  \right).
\end{eqnarray}
Moreover,  if 
\begin{eqnarray}
h^{\mu0}\left(-i D \right)=0, \qquad h^{ij} \left(-i D \right) =-l^{2} D^{i}D^{j},
\end{eqnarray}
we arrive to 
\begin{eqnarray}
S&=&\frac{1}{2}g^{2} \int dx^{d+1} Tr \left( 
 {\cal  F}_{\mu\nu}  {\cal F}^{\mu \nu}-l^{2} {\cal F}_{i}\-^{k} D^{i}D^{j} {\cal F}_{j k}  \right)\nonumber\\
 &=&\frac{1}{2}g^{2}  \int dx^{d+1} Tr \left( 
 {\cal  F}_{\mu\nu}  {\cal F}^{\mu \nu}+l^{2} D_{i}{\cal F}_{ik} D_{j} {\cal F}_{j k}  \right),
 \end{eqnarray}
which is the action for the anisotropic Yang-Mills  with dynamic exponent  $z=2,$ \cite{anselmi6:gnus}. 
Therefore,  this anisotropic Yang-Mills theory  can be seen as a Yang-Mills theory in a momentum dependent metric. 
\section{Noncommutative case}

In this section we will consider  the Hamiltonian action 
\begin{eqnarray}
S=\int d\tau \left( - \Lambda^{\mu \nu}(p)  \dot p_{\mu} x_{\nu} -H(x,p)    \right). \label{eq:nca}
\end{eqnarray}
This  action implies the equations of motion 
\begin{eqnarray}
\dot x^{\mu}&=& \left( \Lambda ^{-1} \right)  ^{\mu \alpha} \left (  \frac{ \partial \Lambda_{\alpha \rho  }} {\partial  p ^{\nu}}  - 
 \frac{ \partial \Lambda_{\nu \rho  }} {\partial  p ^{\alpha}} 
 \right ) x^{\rho} \left( \Lambda^{-1} \right)  ^{\nu \gamma} \frac{\partial H}{\partial x^{\gamma} }+  \left( \Lambda ^{-1} \right)  ^{\mu \gamma} \frac{\partial H}{\partial p^{\gamma} }, 
 \label{eq:1} \\
\dot p^{\mu} &=&- \left(\Lambda ^{-1}\right)^{\mu \alpha} \frac{\partial H}{\partial x^{\alpha} }. \label{eq:2}
\end{eqnarray}
It can be shown that these equations of motion are consistent with the symplectic structure 
\begin{eqnarray}
\left \{ x^{\mu}, x^{\nu} \right \} &=& \left( \Lambda ^{-1} \right)  ^{\mu \alpha} \left (  \frac{ \partial \Lambda_{\alpha \rho  }} {\partial  p ^{\beta }}  - 
 \frac{ \partial \Lambda_{\beta \rho  }} {\partial  p ^{\alpha}} 
 \right ) x^{\rho} \left( \Lambda ^{-1} \right)  ^{\beta \nu}, \label{eq:3}\\
\left\{ x^{\mu}, p^{\nu} \right\}&=&\left(\Lambda ^{-1}\right)^{\mu \nu },  \label{eq:4}\\
\left\{ p^{\mu}, p^{\nu} \right\}&=&0.\label{eq:5}
\end{eqnarray}
Notice that the quantum version of this symplectic structure  is  a noncommutative space-time. \\

Now, we can see that  if  $H$ does  not  depend on $x,$ the equations of motion (\ref{eq:1})-(\ref{eq:2}) are
\begin{eqnarray}
\dot x_{\mu}&=&   \left( \Lambda ^{-1} \right)  ^{\mu \gamma} \frac{\partial H}{\partial p^{\gamma} },  \\
\dot p_{\mu} &=&0, 
\end{eqnarray}
namely,
\begin{eqnarray}
\ddot x_{\mu}&=& 0. 
\end{eqnarray}
Then, in all noncommutative space-time, the classical free particle follows the standard dynamics. \\

We can see  that if  the coordinate $\left(\tilde x^{\mu}, \tilde p_{\mu}\right)$ satisfy the usual Poisson brackets, then
the phase space coordinates 
\begin{eqnarray}
x^{\mu}=\left(\Lambda ^{-1}\right)^{\mu\nu} \tilde x_{\nu}, \qquad \tilde p_{\mu}=p_{\mu},  \label{darboux}
\end{eqnarray}
satisfy the relations (\ref{eq:3})-(\ref{eq:5}), this is the so-called Darboux mapping, which is only locally defined. \\

If we take the modified action for the free relativistic particle 
\begin{eqnarray}
S=\int d\tau \left( - \Lambda ^{\mu \nu}(p)  \dot p_{\mu} x_{\nu} - \frac{ \lambda}{2} \left ( \eta^{\mu\nu}  p_{\mu} p_{\nu} -m^{2} \right)\right),
\end{eqnarray}
we obtain the usual  constraint
\begin{eqnarray}
 p^{2} -m^{2}\approx 0. \label{constrain}
\end{eqnarray}
Notice that we are working  in the extended phase space \cite{dirac1:gnus} where all the momenta are independent and are not restricted by the constraint (\ref{constrain}). In this way, the symplectic structure will be invertible.\\

When in a noncommutative space-time there are   no constant parameters in its commutation relations,  
construct an interacting field theory in that  space-time is a very difficult task. Some advances in this topic can be seen in \cite{kontsevich} and some  
proposals  to obtain  a  field theory in the noncommutative Snyder space-time can be seen in \cite{snyder4:gnus,snyder7:gnus}. 
Construct a field theory in a  noncommutative space-time is not the main issue of this paper.  However,  in different noncommutative space-times, the free particle is a special case. In fact, in some noncommutative space-times the free scalar field is not different form the free scalar field in a commutative space-time \cite{szabo:gnus,sz2:gnus}. 
Notice that using  local Darboux  coordinates (\ref{darboux}), the equation
\begin{eqnarray}
\left( -  \tilde \partial^{\mu} \tilde \partial _{\mu} -m^{2}\right) \phi =0
\end{eqnarray}
can be proposed as a quantum version of the constraints (\ref{constrain}). In the next section we will  propose different cases of $\Lambda^{\mu\nu}(p).$

\subsection{Snyder space-time}

Snyder space-time is an important example of a noncommutative space-time.  For this case, if we take
the following matrix
\begin{eqnarray}
\left ( \Lambda^{-1} \right) ^{\mu\nu}(p ) = \eta^{\mu \nu} + a^{2} p^{\mu}p^{\nu}, \qquad a={\rm constant},
\end{eqnarray}
we obtain the symplectic structure 
\begin{eqnarray}
\left \{ x^{\mu}, x^{\nu} \right \} &=& a^{2}  \left( x^{\mu} p^{\nu} - x^{\nu} p^{\mu} \right), \label{eq:s1}\\
\left\{ x^{\mu}, p^{\nu} \right\}&=&\eta^{\mu \nu} + a^{2} p^{\mu}p^{\nu},\label{eq:s2}\\
\left\{ p^{\mu}, p^{\nu} \right\}&=&0.\label{eq:s3}
\end{eqnarray}
This symplectic structure is compatible with the noncommutative Snyder space-time \cite{s:gnus}:
\begin{eqnarray}
\left [ \hat x^{\mu}, \hat x^{\nu} \right ] &=&i a^{2}  \left( \hat x^{\mu} \hat p^{\nu} - \hat x^{\nu} \hat p^{\mu} \right), \\
\left[ \hat x^{\mu}, \hat p^{\mu} \right]&=&i \left( \eta^{\mu \nu} + a^{2} \hat p^{\mu}\hat p^{\nu}\right),\\
\left[ \hat p^{\mu}, \hat p^{\mu} \right]&=&0.
\end{eqnarray}

In general, if $f(p^{2}) $ is a smooth function, we can propose  
\begin{eqnarray}
\left ( \Lambda^{-1} \right) ^{\mu\nu}(p ) = \eta^{\mu \nu} + f(p^{2}) p^{\mu}p^{\nu},
\end{eqnarray}
which implies the symplectic structure 
\begin{eqnarray}
\left \{ x^{\mu}, x^{\nu} \right \} &=& f(p^{2})  \left( x^{\mu} p^{\nu} - x^{\nu} p^{\mu} \right),  \label{eq:gs1}\\
\left\{ x^{\mu}, p^{\nu} \right\}&=&\eta^{\mu \nu} + f(p^{2}) p^{\mu}p^{\nu},\label{eq:gs2}\\
\left\{ p^{\mu}, p^{\nu} \right\}&=&0.\label{eq:gs3}
\end{eqnarray}
The quantum version of these last Poisson brackets imply  a noncommutative space-time.

\subsection{Euclidean Snyder space-time }

Furthermore,   if we  take 
\begin{eqnarray}
\left( \Lambda^{-1} \right) ^{00} =-1, \quad  \left(\Lambda^{-1}\right) ^{0i}=0, \quad  \left(\Lambda^{-1}\right) ^{ij}= \delta^{ij} +a^{2} p^{i}p^{j},
\end{eqnarray}
we have the following  Poisson brackets:
\begin{eqnarray}
\left \{ x^{0}, x^{\nu} \right \} &=&0,\\
\left \{ x^{i}, x^{j} \right \} &=&a^{2} \left( x^{i} p^{j} - x^{j} p^{i} \right), \\
\left\{ x^{0}, p^{\mu} \right\}&=&\eta^{0 \mu} ,\\
\left\{ x^{i}, p^{j} \right\}&=&\delta^{ij} + a^{2}p^{i}p^{j},\\
\left\{ p^{\mu}, p^{\nu} \right\}&=&0.
\end{eqnarray}
It can be show that the quantum version of this space-time implies discrete geo\-metric  quantities \cite{za:gnus}.\\

In general,  if  $f(\vec p^{\ 2})$ is a smooth function, we can propose the matrix  
\begin{eqnarray}
\left( \Lambda^{-1} \right) ^{00} =-1, \quad  \left(\Lambda^{-1}\right) ^{0i}=0, \quad  \left(\Lambda^{-1}\right) ^{ij}= \delta^{ij} +f(\vec p^{\ 2}) p^{i}p^{j},
\end{eqnarray}
which implies the Poisson brackets 
\begin{eqnarray}
\left \{ x^{0}, x^{\nu} \right \} &=&0,\label{eq:es1}\\
\left \{ x^{i}, x^{j} \right \} &=& f(\vec p^{\ 2})  \left( x^{i} p^{j} - x^{j} p^{i} \right), \\
\left\{ x^{0}, p^{\mu} \right\}&=&\eta^{0 \mu} ,\\
\left\{ x^{i}, p^{j} \right\}&=&\delta^{ij} + f(\vec p^{\ 2}) p^{i}p^{j},\\
\left\{ p^{\mu}, p^{\mu} \right\}&=&0.\label{eq:es2}
\end{eqnarray}
The quantum version of these Poisson brackets implies  a noncommutative space-time.

\section{Noncommutative space-time and Lifshitz field theory}

The modified free particle  given by
\begin{eqnarray}
S=\int d\tau \left( - \Lambda^{\mu \nu}(p)  \dot p_{\mu} x_{\nu} - \frac{ \lambda}{2} \left ( \Omega^{\mu\nu}(p)  p_{\mu} p_{\nu} -m^{2} \right)\right)
\end{eqnarray}
includes both modified free particles studied before. For this system we have the constraint 
\begin{eqnarray}
 \Omega^{\mu\nu}(p)  p_{\mu} p_{\nu} -m^{2}\approx 0.\label{eq:slc1}
\end{eqnarray}
and  the symplectic structure (\ref{eq:3})-(\ref{eq:5}). Notice that the classical equations of motion for this system are   the classical equations of motion for the usual classical relativistic free particle. However, in suitable local Darboux coordinates  (\ref{darboux}), we get the wave equation
\begin{eqnarray}
\left ( -\Omega^{\mu\nu}\left (-i \tilde \partial \right) \tilde  \partial _{\mu} \tilde \partial _{\nu} -m^{2}\right) \phi =0, \label{eq:mass}
\end{eqnarray}
as  the quantum version of the constraint  (\ref{eq:slc1}). This last equation  is a  modified Klein-Gordon equation. 
Notice that   the quantum propagation of this free particle is  changed.\\

We do not know how  an interacting field theory  in this new framework is. However,  in the limit 
\begin{eqnarray}
\Lambda^{\mu\nu}(p) \to \eta^{\mu\nu},
\end{eqnarray}
for non trivial $\Omega^{\mu\nu}(p),$  we should obtain  a Lifshitz-like field theory.  In addition, in the limit
\begin{eqnarray}
\Omega^{\mu\nu}(p) \to \eta^{\mu\nu}
\end{eqnarray}
we should obtain  an usual interacting field theory in a noncommutative space-time.  \\

In the usual particle,  the Minkowski metric appears in the term $\eta^{\mu\nu}\dot p_{\mu}x_{\nu}$ and in the term $\eta^{\mu\nu} p_{\mu}p_{\nu}.$ 
Then in this case we have
\begin{eqnarray}
 \Omega^{\mu\nu}(p)  =\eta^{\mu\nu}= \Lambda^{\mu \nu}(p).
 \end{eqnarray}
For this  reason, we can take the modified action 
\begin{eqnarray}
S=\int d\tau \left( - \Omega^{\mu \nu}(p)  \dot p_{\mu} x_{\nu} - \frac{ \lambda}{2} \left ( \Omega^{\mu\nu}(p)  p_{\mu} p_{\nu} -m^{2} \right)\right)
\end{eqnarray}
as a special case. This last system is interesting, because a Lifshitz field theory is related with a  noncommutative space-time. In the next subsection we 
study some cases.

\subsection{Snyder space-time}

If we take the matrix 
\begin{eqnarray}
\left ( \Omega^{-1} \right) ^{\mu\nu}(p )=\left ( \Lambda^{-1} \right) ^{\mu\nu}(p ) = \eta^{\mu \nu} + l^{2} p^{\mu}p^{\nu}, 
\end{eqnarray}
we obtain the Snyder noncommutative space-time (\ref{eq:s1})-(\ref{eq:s3}) and the constraint (\ref{eq:slc1}) is given by 
\begin{eqnarray}
p^{2}  -m^{2}- \frac{l^{2}}{ 1+l^{2} p^{2} } (p^{2})^{2} =0.
\end{eqnarray}
At first order, this constraint can be written as 
\begin{eqnarray}
p^{2}  -m^{2}- l^{2} (p^{2})^{2} =0.
\end{eqnarray}
Then, at this order, the wave equation for this system is given by 
\begin{eqnarray}
\left( - \tilde \partial_{\mu} \tilde \partial^{\mu} -m^{2}- l^{2}  ( \tilde \partial^{\mu} \tilde  \partial _{\mu}   )^{2} \right ) \phi =0.
\end{eqnarray}

In general, if $f(p^{2}) $ is a smooth function, we can propose  
\begin{eqnarray}
\left ( \Omega^{-1} \right) ^{\mu\nu}(p )=\left ( \Lambda ^{-1} \right) ^{\mu\nu}(p ) = \eta^{\mu \nu} + f(p^{2}) p^{\mu}p^{\nu},
\end{eqnarray}
which implies the symplectic structure (\ref{eq:gs1})-(\ref{eq:gs3}). For this case, the constraint (\ref{eq:slc1}) is given by 
\begin{eqnarray}
p^{2}  -m^{2}- \frac{f(p^{2})}{ 1+f(p^{2}) p^{2} } (p^{2})^{2} =0.
\end{eqnarray}
At first order, this constraint can be written as 
\begin{eqnarray}
p^{2}  -m^{2}- f(p^{2}) (p^{2})^{2} =0.
\end{eqnarray}
At quantum level, this constraint implies  the wave equation 
\begin{eqnarray}
\left( - \tilde \partial_{\mu} \tilde \partial^{\mu} -m^{2}- f\left (-\tilde \Box  \right) (\tilde  \partial^{\mu}\tilde  \partial _{\mu} )^{2} \right) \phi =0.
\end{eqnarray}

\subsection{Euclidean Snyder space-time and  Lifshitz field theory}

Now, if we take 
\begin{eqnarray}
& &\left( \Omega^{-1} \right) ^{00} =\left( \Lambda^{-1} \right) ^{00} =-1, \quad  \left(\Omega ^{-1}\right) ^{0i}=\left(\Lambda ^{-1}\right) ^{0i}=0, \nonumber\\
 & &\left(\Omega^{-1}\right) ^{ij}=\left(\Lambda^{-1}\right) ^{ij}= \delta^{ij} +f(\vec p^{\ 2}) p^{i}p^{j}, \nonumber
\end{eqnarray}
we have the   Poisson brackets (\ref{eq:es1})-(\ref{eq:es2})  and the constraint 
\begin{eqnarray}
p^{2}  -m^{2}- \frac{f(\vec p^{\ 2})}{ 1+f(\vec p^{\ 2}) \vec p^{\ 2} } (\vec p^{\ 2})^{2} =0.
\end{eqnarray}
At first order, this constraint can be written as 
\begin{eqnarray}
p^{2}  -m^{2}- f(\vec p^{\ 2}) (\vec p^{\ 2})^{2} =0.
\end{eqnarray}
Then, at this order, the wave equation for this system is given by 
\begin{eqnarray}
\left( - \tilde \partial_{\mu} \tilde \partial^{\mu} -m^{2}- f\left (- \tilde \nabla^{2}\right ) (\tilde \nabla^{2} )^{2} \right) \phi =0.\label{eq:lifshitz}
\end{eqnarray}
In particular, if 
\begin{eqnarray}
f(\vec p^{\ 2}) = a (-)^{z-1} (\vec p^{\ 2})^{z-2},
\end{eqnarray}
the wave equation (\ref{eq:lifshitz}) becomes 
\begin{eqnarray}
\left( - \tilde \partial_{\mu} \tilde \partial^{\mu} -m^{2}- a ( \tilde \nabla^{2})^{z}  \right) \phi =0,
\end{eqnarray}
which is the equation of motion for a scalar Lifshitz field theory \cite{anselmi10:gnus}.
In addition, this system lives in the following noncommutative space-time:
\begin{eqnarray}
 \left[ \hat x^{0}, \hat x^{\nu} \right ] &=&0,\\
\left [\hat x^{i}, \hat x^{j} \right ] &=& ia (-)^{z-1} \left (\hat  p^{2}\right)^{z-2} \left( \hat x^{i} \hat p^{j} - \hat x^{j} \hat p^{i} \right), \\
\left[ \hat x^{0}, \hat p^{\mu} \right]&=&\eta^{0 \mu} ,\\
\left[ \hat x^{i}, \hat p^{j} \right]&=&i\delta^{ij} + ia (-)^{z-1} \left(\hat  p^{2}\right)^{z-2} \hat p^{i}\hat p^{j},\\
\left[ \hat p^{\mu}, \hat p^{\mu} \right]&=&0.
\end{eqnarray}
Notice that when $z=2,$ this noncommutative space-time implies discrete geometric quantities \cite{za:gnus}.

\section{Summary}
In this work, we proposed three different modified first order actions for relativistic particles. 
In the first case we  proposed a particle with a momentum dependent metric and 
we showed  that  the quantum version of these systems  include  different field theories,  as Lifshitz  field theories. 
As a second case we proposed a particle  that implies a modified symplectic structure and we showed   that the quantum version of this system gives  different noncommutative space-times, for example the Snyder space-time. In the third  case, we combined both structures  before mentioned, namely  noncommutative space-times and momentum dependent metric. In this last case, we showed  that anisotropic field theories can be seen as a limit of noncommutative field theory. 

\section{Acknowledgments }
This work was supported in part by DGAPA grant IN109013 (J.D.V.) and CONACyT-SEP 47510318 (J.M.R.).

\end{document}